\documentclass[preprint,12pt]{elsarticle}

\journal{Parallel Computing}

\usepackage{subfigure}
\usepackage{graphicx}
\usepackage{amsmath}
\usepackage{amsthm}
\usepackage{amsfonts}
\usepackage{enumerate}
\usepackage[boxed,lined]{algorithm2e}

\begin{document}

\begin{frontmatter}




\title{
Batched Kronecker product for 2-D matrices and 3-D arrays on NVIDIA GPUs
}


\author{Chetan Jhurani}
\ead{chetan.jhurani@gmail.com}

\address{
	Tech-X Corporation\\
	5621 Arapahoe Ave\\
	Boulder, Colorado 80303, U.S.A.
}


\begin{abstract}

We describe an interface and an implementation for performing
Kronecker product actions on NVIDIA GPUs for multiple small 2-D
matrices and 3-D arrays processed in parallel as a batch.  This method
is suited to cases where the Kronecker product component matrices are
identical but the operands in a matrix-free application
vary in the batch.  Any batched GEMM (General Matrix Multiply) implementation,
for example ours~\cite{tgemm} or the one in cuBLAS~\cite{cublas},
can also be used for performing batched Kronecker
products on GPUs.  However, the specialized implementation presented
here is faster and uses less memory.  Partly this is because a simple
GEMM based approach would require extra copies to and from main
memory.  We focus on matrix sizes less than or equal to
16, since these are the typical polynomial degrees in Finite Elements,
but the implementation can be easily extended for other sizes. 
We obtain 143 and 285 GFlop/s for single precision real when
processing matrices of size 10 and 16, respectively on NVIDIA Tesla
K20c using CUDA 5.0.  The corresponding speeds for 3-D array Kronecker products are
126 and 268 GFlop/s, respectively.
Double precision is easily supported using the C++ template mechanism.

\end{abstract}


\end{frontmatter}



\newcommand{\matspace}[3]{\ensuremath{\mathbb{#1}^{#2 \times #3}} }

\newcommand{\Rnn}  {\matspace{\mathbb{R}}{n}{n}}
\newcommand{\Rmn}  {\matspace{\mathbb{R}}{m}{n}}
\newcommand{\Rmm}  {\matspace{\mathbb{R}}{m}{m}}

\newcommand{\Cnn}  {\matspace{\mathbb{C}}{n}{n}}
\newcommand{\Cmn}  {\matspace{\mathbb{C}}{m}{n}}
\newcommand{\Cmm}  {\matspace{\mathbb{C}}{m}{m}}
\newcommand{\Crn}  {\matspace{\mathbb{C}}{r}{n}}
\newcommand{\Crr}  {\matspace{\mathbb{C}}{r}{r}}


\newcommand{\vecspace}[2]{\ensuremath{\mathbb{#1}^{#2}} }

\newcommand{\Rn}   {\vecspace{\mathbb{R}}{n}}
\newcommand{\Cn}   {\vecspace{\mathbb{C}}{n}}

\newcommand{\Rm}   {\vecspace{\mathbb{R}}{m}}
\newcommand{\Cm}   {\vecspace{\mathbb{C}}{m}}

\newcommand{\reals} {\ensuremath{\mathbb{R}}}
\newcommand{\complex} {\ensuremath{\mathbb{C}}}


\newcommand{\norm}[1]{\ensuremath{\left| \left| #1 \right| \right|} }

\newcommand{\abs}[1]{\ensuremath{\left| #1 \right|} }

\newcommand{\pinv}[1]{\ensuremath{{#1}^{\dagger}} }
\newcommand{\inv}[1]{\ensuremath{{#1}^{-1}} }
\newcommand{\h}[1]{\ensuremath{{#1}^{op}} }

\newcommand{\nullsp}[1]{\ensuremath{\mathcal{N}(#1)} }
\newcommand{\rangsp}[1]{\ensuremath{\mathcal{R}(#1)} }

\newcommand{\patsym}[0]{\ensuremath{\mathcal{Z}} }
\newcommand{\pat}[1]{\ensuremath{\mathcal{Z}(#1)} }

\newcommand{\lagr}[0]{\ensuremath{\mathcal{L}} }

\newcommand{\partderiv}[2]{\ensuremath{\frac{\partial #1}{\partial #2}} }

\newcommand{\half}[0]{\ensuremath{\frac{1}{2}} }


\newcommand{\ordset}[3] {\ensuremath{\left\{ {#1}_{#2} \right\}_{#2 = 1}^{#3}}}


\newcommand{\Eq}[1]  {Equation~(\ref{#1})}
\newcommand{\Eqs}[1] {Equations~(\ref{#1})}
\newcommand{\Sec}[1] {Section~\ref{#1}}
\newcommand{\Fig}[1] {Figure~\ref{#1}}
\newcommand{\Alg}[1] {Algorithm~\ref{#1}}
\newcommand{\Rem}[1] {Remark~\ref{#1}}
\newcommand{\Prop}[1] {Property~\ref{#1}}
\newcommand{\Thm}[1] {Theorem~\ref{#1}}
\newcommand{\Def}[1] {Definition~\ref{#1}}
\newcommand{\Tab}[1] {Table~\ref{#1}}


\newtheorem{theorem}{Theorem}[section]
\newtheorem{lemma}[theorem]{Lemma}
\newtheorem{proposition}[theorem]{Proposition}
\newtheorem{corollary}[theorem]{Corollary}
\newtheorem{conjecture}[theorem]{Conjecture}
   
\theoremstyle{definition}
\newtheorem{definition}[theorem]{Definition}
\newtheorem{example}[theorem]{Example}
\newtheorem{examples}[theorem]{Examples}

\theoremstyle{remark}
\newtheorem{remark}[theorem]{Remark}



\section{Introduction}

The Kronecker product is an important mathematical concept as well as
a practical tool in multiple applications~\cite{vanloankron}. It is useful where either
the domain or range of a linear operator or the space of unknowns is a
``vector space'' of matrices.  It is commonplace for operator evaluation
in high-order finite elements~\cite{deville2002hom,hackbusch-tensor}, control theory~\cite{cite:lyap2},
and when tensor-like decomposition is possible for different ``directions''~\cite{Loan93approximationwith,hackbusch-tensor}.

Dense Kronecker product action
has an additional advantage of possessing high arithmetic intensity
just like the General Matrix Multiply (GEMM) routine in BLAS~\cite{cite:lapack}.  This is
important on CPUs as well as GPUs~\cite{fermi_gemm}.  In fact, matrix-free
action of a Kronecker product can also be cast as a sequence of GEMM operations.  Thus, modern
computational hardware can achieve high performance in computing
Kronecker product action.  This makes it important to recognize if a computation
can be cast in terms of Kronecker products and utilize it just like
GEMM.

Our goal and contribution is to obtain high performance of Kronecker
product action, possibly even higher than GEMM.  We
emphasize that we never form the explicit product matrix since it is
rarely required.  In our case, the sizes are small (less than or equal
to 16) and the matrices are square but the method can be easily implemented for rectangular cases and for larger sizes.  Such
sizes correspond to the polynomial degrees and quadrature
used in high-order finite elements~\cite{spectralHP,deville2002hom,cite:hpbook,cite:hpbook2,solin2003higher}.
However, we are unaware
of any previous work done on GPUs.  Part of the reason why this is so
can be attributed to a lack of very high performance GEMM for batches
of small matrices, a deficiency that we addressed in~\cite{tgemm}. NVIDIA's
cuBLAS library also contains a batched GEMM implementation~\cite{cublas},
which is different than ours.

If high-performance GEMM for small matrices is available, then one might
deduce that fast Kronecker product action is a solved problem.  This
reasoning is not entirely correct though.  We will show that the
performance of even a fast GEMM implementation is less than a specialized
Kronecker product action.  This can be due to two separate reasons.  In the
first case, some components of the Kronecker product do not change in
the batch and a general GEMM based approach cannot fully take advantage of this.  In
the second case, the domain and range of the action are not (2-D)
matrices but 3-D arrays and more data movement to and from GPU memory reduces
the overall speed.  Both 2-D and 3-D cases arise in practice~\cite{deville2002hom}.

In order to achieve our goal, we also design a general BLAS-like
function interface for Kronecker product action.  Neither BLAS nor
LAPACK implement such an interface~\cite{cite:lapack}, although it is possible to
integrate it easily and would fit perfectly in their design
philosophy.  We design interfaces for matrices as well as 3-D arrays.

We obtain 143 and 285 GFlop/s for single precision real when
processing matrices of size 10 and 16, respectively on NVIDIA Tesla
K20c using CUDA 5.0.  The corresponding speeds for 3-D array Kronecker products are
126 and 268 GFlop/s, respectively.  This performance is appreciably better
than current batched GEMM performance~\cite{tgemm} and justifies the effort
required for a specialized implementation.

Here is an outline of the paper.  In~\Sec{sec:kron} we 
describe our notation for Kronecker products for 2-D as well as 3-D batched
inputs.
\Sec{sec:kron_intf} shows a few function interfaces required
for the implementation. We give
an algorithmic overview of our implementation in~\Sec{sec:kron_impl}.  Finally, we present
the performance of our implementation in \Sec{sec:kron_result} for various
inputs and compare it to the performance of batched GEMM.

\section{Kronecker product in 2-D and its generalizations}
\label{sec:kron}

The standard way of defining Kronecker product is to define it as an
output matrix that is computed from two arbitrary real or complex
input matrices.  We work in the complex field below for generality.
Let $A \in \complex^{m_a \times n_a}$ and $B \in \complex^{m_b \times
n_b}$.  The Kronecker product of $A$ and $B$ is denoted by $A \otimes
B$. It is an element of $\complex^{m_a m_b \times n_a n_b}$, such that
$$
A \otimes B =
\begin{bmatrix}
	a_{11} B & \cdots & a_{1 n_a} B \\
	\vdots   & \ddots & \vdots   \\
	a_{m_a 1} B & \cdots & a_{m_a n_a} B
\end{bmatrix}
$$
decomposed block-wise.  See~\cite{cite:highamASNA2,hornjohnson} for it properties and applications.

Rather than looking at the Kronecker product as a matrix, we, as is
also common, think of it as a linear operator defined by its two input
matrices $A$ and $B$.  The domain of the operator $A \otimes B$ is
$\complex^{n_b \times n_a}$ and the range is $\complex^{m_b \times
m_a}$.  The operator is defined by its action on a matrix $X \in
\complex^{n_b \times n_a}$.  We have
$$
(A \otimes B) \text{vec}(X) = \text{vec}(B X A^T)
$$
where ``vec'' is the vectorization operation that stacks columns of a
matrix to form a vector, and $Y \in \complex^{m_b \times m_a}$ is the output matrix.
The superscript $T$ stand for transpose without
any conjugation.  One advantage of this is that forming the product
matrix, which can be much larger, is not necessary for it to be
applied.  Another advantage is that computing matrix products using
matrices $A$ and $B$ is faster.  This is due to fewer floating point
operations (by counting flops~\cite{vanloankron}) and because a BLAS level-3 operation
GEMM is used in practice.

\subsection{Generalizations to other dimensions}

We now describe generalizations for the Kronecker product action so that
the input and output need not be restricted to (2-D) matrices.  The
generalized mapping is from $n$-D arrays to $n$-D arrays for $n \geq 1$.  However,
to avoid complicating the notation, we work with $n \leq 3$ only.  One
application where the 3-D case arises as naturally as the 2-D case is
when computing stiffness matrix action on tensor product finite
elements~\cite{deville2002hom}.

First, let us unravel the matrix symbolism present in the 2-D case so
that the summations are visible.  We change our formulation a little
so that the matrix $A$ is related to the first direction in $X$
and the matrix $B$ is related to the second direction in $X$.
This will make the generalization natural when we proceed to 3-D.
We now work with
the operator $B \otimes A$ in the 2-D case, where $A$ and $B$ are of
the same sizes as mentioned above.  But now $X \in \complex^{n_a
\times n_b}$ and $Y = A X B^T \in \complex^{m_a \times m_b}$.
Expanding this, we get
$$
Y_{ij} = \sum_l^{n_a} \sum_m^{n_b} A_{il} X_{lm} B_{jm} = \sum_l^{n_a} \sum_m^{n_b} A_{il} B_{jm} X_{lm}.
$$
We generalize this summation to 1-D and 3-D, which means the input
``$X$'' and output ``$Y$'' will both be either 1-D or 3-D arrays.
Extension to higher order products will be obvious.

In 1-D, we have a single summation as follows.  Only matrix $A$ is needed.
$$
Y_i = \sum_l^{n_a} A_{il} X_l
$$
This case has a more familiar name.  It is just a simple matrix-vector multiplication,
$Y = A X$, where $Y$ and $X$ are vectors.

In 3-D, we have a triple summation as follows.  This requires
introduction of a matrix $C \in \complex^{m_c \times n_c}$.  We have
$X \in \complex^{n_a \times n_b \times n_c}$ and $Y \in
\complex^{m_a \times m_b \times m_c}$, where
$$
Y_{ijk} = \sum_l^{n_a} \sum_m^{n_b} \sum_n^{n_c} A_{il} B_{jm} C_{kn} X_{lmn}.
$$
It can be shown that the equation above is an expanded form of the
equation below.
$$
\text{vec}(Y) = (C \otimes B \otimes A) \text{vec}(X)
$$
Here ``vec'' vectorizes 3-D arrays ($X$ and $Y$) similar to the 2-D
column major order vectorization.  The order of Kronecker operations
in $C \otimes B \otimes A$ is immaterial and thus Kronecker product is
associative.

We hide the vectorization operation notation below by denoting
the 1-D operation by $Y = \mathcal{K}(A)(X)$, the 2-D
operation by $Y = \mathcal{K}(A,B)(X)$, and the 3-D
operation by $Y = \mathcal{K}(A,B,C)(X)$.  This emphasizes
that we are concerned with a linear operation on $n$-D arrays
rather than computing a Kronecker product matrix.

\subsection{Batched Kronecker products}

We now describe the batched Kronecker product operation as it needs to
be implemented in practice.  Our choices are inspired by the BLAS/LAPACK
interface design~\cite{cite:lapack}.

Let $op$ (standing for operation) refer to a mapping from matrices to
matrices, such that $op(A) = A$ or $A^T$ or $A^*$ depending on an
extra variable that can contain three values.  The superscripts $T$
and $*$ stand for transpose and Hermitian-transpose, respectively.
Fix real or complex matrices $A, B,$ and $C$.  The matrices can be stored in
single precision or double precision.  Using the BLAS notation,
$op(A)$ is $m_a \times n_a$, and $op(B)$ is $m_b \times n_b$, and
$op(C)$ is $m_c \times n_c$. 

Consider the 3-D case first.  The inputs are $X^p$ for $p =
1,2,\ldots,N$.  Here $N$ stands for the batch size. Each $X$ is $n_a
\times n_b \times n_c$.  The outputs are $Y^p$, each being $m_a \times
m_b \times m_c$, and are computed as follows.
$$
Y^p \leftarrow \alpha\, \mathcal{K}(op(A),op(B),op(C))(X^p) + \beta\, Y^p.
$$
Here $\alpha$ and $\beta$ are given scalar values.  Note that the
matrices $A$, $B$, and $C$ do not change in the batch.

The 2-D case is almost the same, except that it is meaningful to have
the $op$ operation apply on $X$ as well.  We then have each $op(X)$ of
size $n_a \times n_b$. The outputs are $Y^p$, each being $m_a \times
m_b$.  We have
$$
Y^p \leftarrow \alpha\, \mathcal{K}(op(A),op(B))(op(X^p)) + \beta\, Y^p.
$$

The operations $op$ acting on $A$, $B$, and $C$ can be different,
which leads to $3^3 = 27$ combinations in all for the complex case and
$2^3 = 8$ combinations for the real case in 3-D.  Of course,
implementing all these cases will lead to code bloat, in which case
one might avoid the full generalization and implement only the specific
cases one is interested in.  Another possibility is specific to our
case.  We have $A$, $B$, and $C$ as constants and thus the cost of
transposing them once before using them for all matrices in the batch
is relatively minor.  This avoids all the $op$ related permutations (except those for
$X$ in the 2-D case) in the implementation.

\section{Batched function interface}
\label{sec:kron_intf}

We present BLAS-like function interfaces for 1-D, 2-D, and 3-D
batched Kronecker products. Each input $X$ and each output $Y$ is
stored in column-major order, just like the matrices $A,B,$ and $C$.
Most of the naming convention below would be natural to BLAS and
cuBLAS~\cite{cublas} users and is easily understood by reading the material
presented earlier.  Thus, we describe only what is new.

The letter {\tt T} below stands for type and is for single or double
precision real or complex.  The two variables that need a little
explanation are {\tt ldx2} and {\tt ldxp} (and other similar ones for
other matrices).  Here {\tt ldx2} denotes the offset in number of
elements between adjacent ``planes'' in the 3-D array $X$.  It is the
the number of elements that are stored in the half-open memory range
corresponding to $[X_{lm1},X_{lm2})$, for example.  Similarly {\tt ldxp}
denotes the offset between adjacent $X$ inputs.  This also
means that each input and output in a batch is stored at uniform
offset from previous entity.  These concepts and arguments for and
against them were discussed in detail in~\cite{tgemm}.  Additionally,
as discussed there, many code optimizations are possible using C++
templates (instead of duplicating code).  To avoid losing focus, we do
not repeat here how C++ templates are used.

Here is the interface for 1-D.  This is almost as if we have
implemented a batched GEMV operation.  Note that having the option for
$op$ on $X$ in 1-D is just an unnecessary complication.  Hence we drop
it.
\begin{verbatim}
void TKRON1(
    char transa,
    int ma, int na,
    int batch_count,
    const T* alpha,
    const T* A, int lda,
    const T* X, int ldp,
    const T* beta,
    T* Y, int ldp);
\end{verbatim}

Here is the interface for 2-D.
\begin{verbatim}
void TKRON2(
    char transa, char transb, char transx,
    int ma, int na,
    int mb, int nb,
    int batch_count,
    const T* alpha,
    const T* A, int lda,
    const T* B, int ldb,
    const T* X, int ldx, int ldxp,
    const T* beta,
    T* Y, int ldy, int ldyp);
\end{verbatim}

Finally, here is the interface for 3-D.  Note that there is no {\tt
transx} input argument.  We are not aware of an application that would
require it.  Besides, generalizing transpose for 3-D array cannot
just be flipping two indices.
\begin{verbatim}
void TKRON3(
    char transa, char transb, char transc,
    int ma, int na,
    int mb, int nb,
    int mc, int nc,
    int batch_count,
    const T* alpha,
    const T* A, int lda,
    const T* B, int ldb,
    const T* C, int ldc,
    const T* X, int ldx, int ldx2, int ldxp,
    const T* beta,
    T* Y, int ldy, int ldy2, int ldyp);
\end{verbatim}

We now give an interface for performing a specialized batched GEMM
operation where the first matrix $A$ varies across the batch but the
second matrix $B$ does not.  This will be used for implementing {\tt
TKRON3} (see \Alg{alg:kron3} ahead). Mathematically, our operation looks like this. Let $\alpha$
and $\beta$ be given scalars and $p = 1,2,\ldots,N$.  We want to
compute
\begin{equation}
C^p \leftarrow \alpha \, op(A^p) op(B) + \beta \, C^p
\end{equation}
using the following interface.  The suffix {\tt A} specifies
that the matrix $A$ varies in the batch and thus requires
a parameter {\tt lda2}.
\begin{verbatim}
void TGEMM_A(
    char transa, char transb,
    int m, int n, int k,
    const T* alpha,
    const T* A, int lda, int lda2,
    const T* B, int ldb,
    const T* beta,
    T* C, int ldc, int ldc2,
    int batch_count,
    int grid_size);
\end{verbatim}

\section{CUDA kernel implementation overview}
\label{sec:kron_impl}

Before describing a few details relevant to the CUDA based
implementation, it is useful to see how the 3-D kernel
{\tt TKRON3} can be implemented in terms of {\tt TKRON2}
and the batched {\tt TGEMM\_A} shown earlier.  To keep
the discussion simpler, we show
this for a non-batched operation, where only one $X$
is an input and only one $Y$ is the output.  This
is shown in \Alg{alg:kron3} using MATLAB notation.

\begin{algorithm}
\label{alg:kron3}

\KwData{$A \in \complex^{m_a \times n_a}$, $B \in \complex^{m_b \times n_b}$, $C \in \complex^{m_c \times n_c}$, $X \in \complex^{n_a \times n_b \times n_c}$}
\vspace{0.4cm}
\KwResult{$Y \in \complex^{m_a \times m_b \times m_c}$ where $\text{vec}(Y) = (C \otimes B \otimes A) \text{vec}(X)$}\vspace{0.5cm}

$Y   \gets \text{zeros}(m_a,m_b,m_c)$\\
$tmp \gets \text{zeros}(m_a,m_b,n_c)$\\
\vspace{0.4cm}
\CommentSty{\% First kron(B,A) is appplied to matrices `in' X.\\}
\CommentSty{\% GEMM operations are used.\\}
\CommentSty{\% This step corresponds to TKRON2.\\}
\vspace{0.4cm}
\For{N = 1 $\to$ $n_c$}{
    $tmp(:,:,N) \gets A * X(:,:,N) * B'$\\
}
\vspace{0.4cm}
\CommentSty{\% Then C is applied to compute slices of Y.\\}
\CommentSty{\% GEMM operations are used again.\\}
\CommentSty{\% reshape is an inexpensive MATLAB function.\\}
\CommentSty{\% This step corresponds to TGEMM\_A.\\}
\vspace{0.4cm}
\For{J = 1 $\to$ $m_b$}{
    $Y(:,J,:) \gets \text{reshape}(\text{reshape}(tmp(:,J,:),m_a,n_c) * C',m_a,1,m_c)$\\
}
\vspace{0.4cm}
\caption{A sequence of GEMM operations in MATLAB notation to compute Kronecker product action on a 3-D array.  We implement this
in CUDA for a batch of inputs in the {\tt TKRON3} interface.}
\end{algorithm}

We need two CUDA kernel implementations --
one for the {\tt TKRON2} operation and the other one for {\tt TGEMM\_A}.
Both kernel implementations resemble the {\tt TGEMM\_multi\_uniform\_kernel}
implementation we discussed in~\cite{tgemm}.
Similar to that case, we have two distinct methods.  The first
one is for matrix sizes 1-16. The second can be used for
matrices where the square matrix dimension can be factorized
into a product of two nearly equal numbers.  For example, 15 can
be factorized as $3 \times 5$ or $5 \times 3$.  The order is
important.  Experimentally, we saw that the second method
is faster than the first one for sizes 15 and 16 and we use that
to show the results. In both methods, each CUDA thread-block is
used to process multiple matrices.

What is more important and different here is that all threads read the ``constant''
matrices in the batch and save them in the shared memory.  
This is an obvious enhancement that helps
in getting a better performance compared to GEMM.
Then, each CUDA block collectively and reads a new $X$ matrix and writes the results
to a single $Y$ matrix.  

Another issue is that (at least) our implementation of these
interfaces requires temporary storage, or ``work arrays''
in the Fortran and LAPACK~\cite{cite:lapack} parlance.  However, we have not explicitly mentioned the details here.  Our description
of the implementation in \Alg{alg:kron3} does show how much temporary storage
is needed in our implementation of {\tt TKRON3}.  We would like to stress here that although
a work array
based interface may look like a hassle and a relic when calling the function, it
has practical and philosophical
advantages even when dynamic allocation facilities are present.  For example, the alternative of memory allocation (and deallocation)
inside the routine might make it more expensive when large temporary storage is
required.  We noticed the detrimental effect on the speed when we did not use work arrays and called {\tt cudaMemalloc} and {\tt cudaFree}
for large temporary storage.
Dynamic allocation also changes the characteristics of a function.  If the function could
be made ``pure'', that opportunity is lost.  One also has to consider multi-threaded
environments, error messages in case of memory errors, multiple devices, and other such responsibilities
that are better handled at some level higher than a high-performance kernel call~\cite{lakos}.
Work arrays provide the full generality with only a minor hassle, and one that can be easily
hidden by writing a small wrapper if one desires.

\section{Performance results}
\label{sec:kron_result}

Our test hardware is the Tesla K20c GPU, which has a peak performance of
3.52 TFlop/s and 1.17 GFlop/s in singe and double precision,
respectively.  We work with the ECC (error-correcting code) mode
turned off for all cases. Our code has been compiled with {\tt
-arch=sm\_35 -O2} options using the CUDA 5.0 toolkit.  To stay within
the global memory limit of the device for the largest matrix size 16, we use 100,000 inputs in the single-precision batch
and half of that in the double-precision batch.  We launch kernels on 5000 CUDA kernel
thread-blocks.  We have not used any transpose or conjugate-transpose operation
in presenting the results, but the actual implementation allows for that.
We have computed all the flop rates by using $4m^3$ and $6m^4$ as the number of 
flops required to perform Kronecker action in 2-D and 3-D, respectively.  This holds
for real matrices that are of size $m$.  We will show results when $\alpha = 1$ and $\beta = 0$.
Finally, the memory layout of matrices is such that all the leading dimension parameters
are the smallest they can logically be for the given matrix sizes.

We show the speeds achievable for real single and double precision data types
in~\Tab{tab:gflop}.  Both 2-D and 3-D results are shown there.  In general,
the 2-D results are slightly faster than the 3-D result.  This is because
the 3-D result requires two sets of kernel calls and the intermediate
result is written to and read from the main memory.  This is the $tmp$
array described in \Alg{alg:kron3}.

\begin{figure}
\centering
\includegraphics[scale=0.4,page=1]{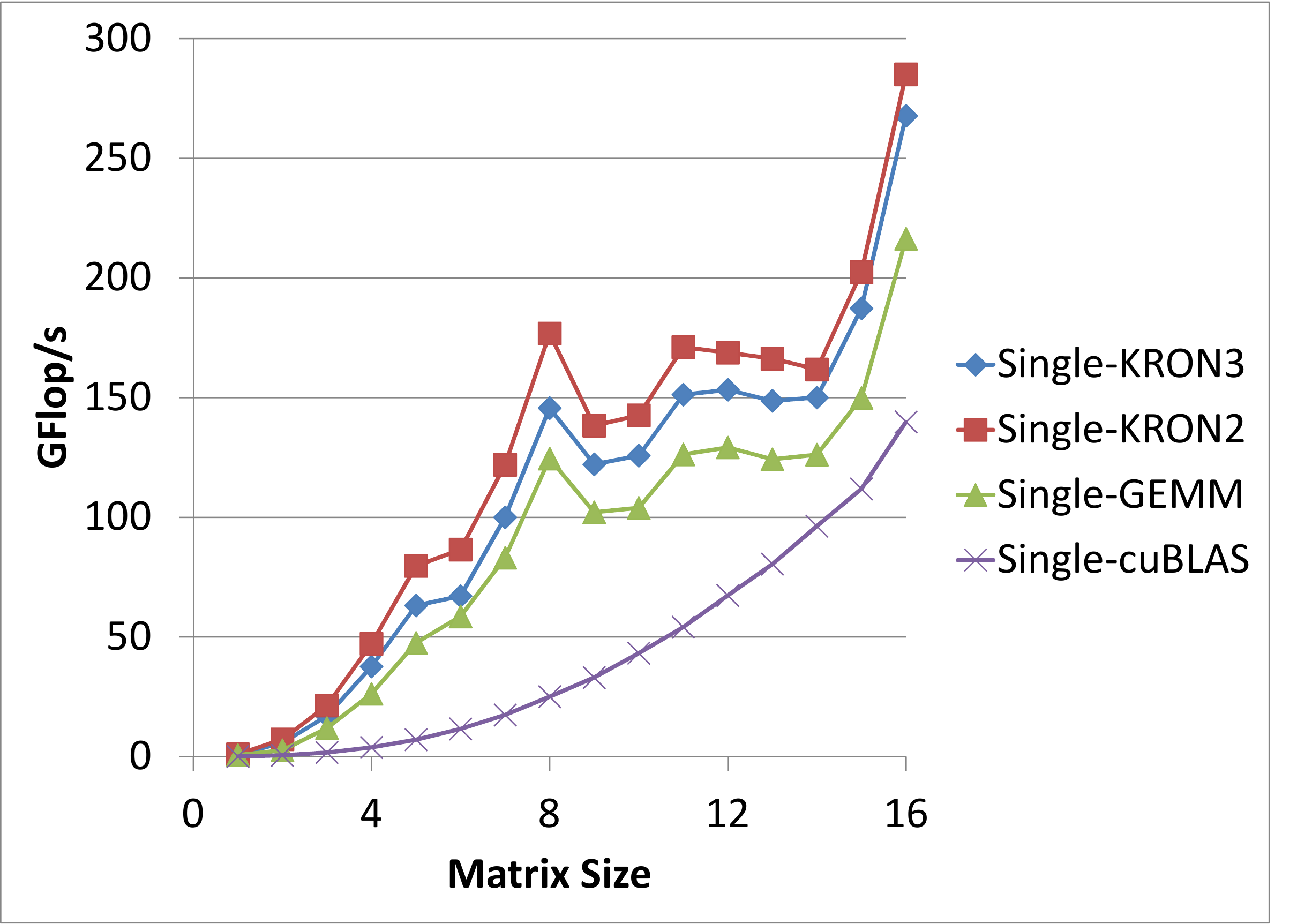}
\caption{Performance obtained for single precision {\tt TKRON2} and {\tt TKRON3} when computing on 100,000
batched inputs with $\alpha = 1$ and $\beta = 0$ on an NVIDIA Tesla K20c.  We show figures for our GEMM from~\cite{tgemm}
for comparison and the batched GEMM implemented in cuBLAS~\cite{cublas}.}
\label{fig:kron_single_compare}
\end{figure}

\begin{figure}
\centering
\includegraphics[scale=0.4,page=2]{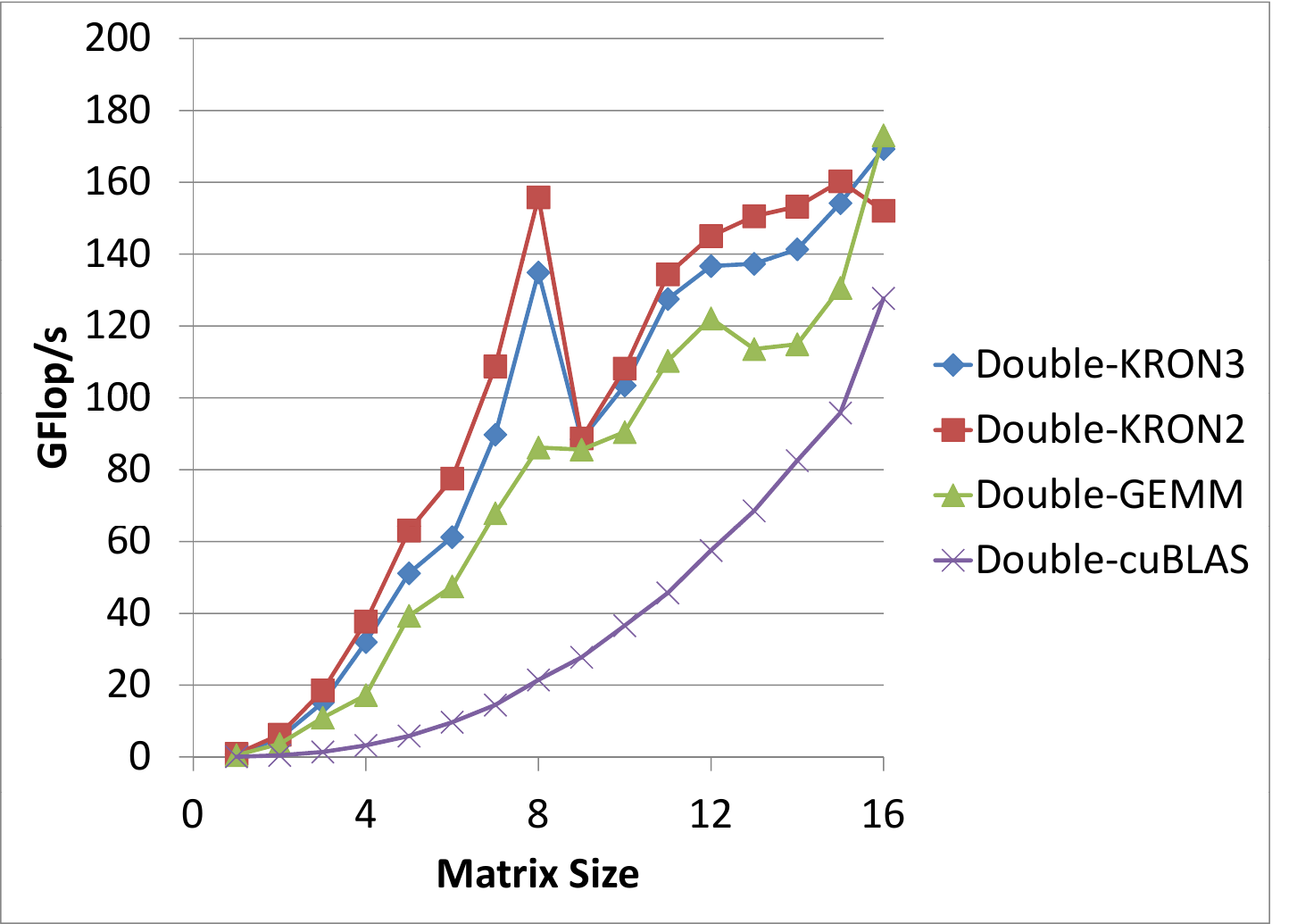}
\caption{Performance obtained for double precision {\tt TKRON2} and {\tt TKRON3} when computing on 50,000
batched inputs with $\alpha = 1$ and $\beta = 0$ on an NVIDIA Tesla K20c.  We show figures for our GEMM from~\cite{tgemm}
for comparison and the batched GEMM implemented in cuBLAS~\cite{cublas}.}
\label{fig:kron_double_compare}
\end{figure}

We also compare the Kronecker product action speeds with the GEMM speeds we
showed in~\cite{tgemm} and the batched GEMM available in cuBLAS.
The results are in \Fig{fig:kron_single_compare} and \Fig{fig:kron_double_compare}
for single and double precision, respectively.  The motivation is to
show that the Kronecker product action can be appreciably faster than
both our GEMM and the one in cuBLAS.  Note that if GEMM were used to
implement Kronecker product, then the effective speed would be even
lower than that of GEMM because of redundant data movement.  The figures
show that the best possible speeds (without such a reduction) are
much lower than our specialized implementation.

\begin{table}
\begin{center}
\begin{tabular}{  r  r  r  r  r }
\hline
Size & Single-2 & Single-3 & Double-2 &  Double-3 \\ \hline
\hline                       
 1 &	1	&	1	&	1	&	1	\\ 
 2 &	7	&	6	&	6	&	5	\\ 
 3 &	21	&	17	&	19	&	15	\\ 
 4 &	47	&	38	&	38	&	32	\\ 
 5 &	80	&	63	&	63	&	51	\\ 
 6 &	86	&	67	&	78	&	61	\\ 
 7 &	122	&	100	&	109	&	90	\\ 
 8 &	177	&	146	&	156	&	135	\\ 
 9 &	138	&	122	&	89	&	88	\\ 
10 &	143	&	126	&	108	&	103	\\ 
11 &	171	&	151	&	134	&	128	\\ 
12 &	169	&	153	&	145	&	137	\\ 
13 &	166	&	149	&	151	&	137	\\ 
14 &	162	&	150	&	153	&	141	\\ 
15 &	202	&	187	&	160	&	154	\\ 
16 &	285	&	268	&	152	&	169	\\ \hline
\end{tabular}
\end{center}
\caption{The GFlop/s rates when computing action of independent Kronecker products of various sizes and precisions
using the {\tt TKRON2} and {\tt TKRON3} interfaces on an NVIDIA Tesla K20c. We use 100,000 inputs for single
precision and 50,000 inputs for double precision (to stay within
global memory limit of the device for the largest matrix size 16).
The comparison with corresponding numbers for a GEMM baseline are made in
\Fig{fig:kron_single_compare} and \Fig{fig:kron_double_compare}.}
\label{tab:gflop}
\end{table}

\section{Discussion}

We have described a BLAS-like interface for a batched Kronecker product
routine generalized to 1-D, 2-D, and 3-D arrays.  The implementation
and results are for 2-D and 3-D only.  Our implementation has assumed that the component matrices
that form the Kronecker product are identical in the batch.  This is
obviously a design choice and based on our application requirement.
If desired, one can similarly implement the computationally more expensive and
general case where one or more of $A$, $B$, or $C$ matrix changes in
the batch.

We have focused on NVIDIA GPUs and used CUDA throughout.  However, the
implementation does not use any special features that might not be
available on other many-core hardware via OpenCL, for example.  A new
implementation will allow devices from other vendors with little or no
change to the interface.


\vspace{1cm}
\noindent \textbf{ \large{Acknowledgments}}

This work was partially supported by the US Department of Energy SBIR Grant
DE-SC0004439.  The author also thanks Paul Mullowney, Tech-X Corporation for
supporting the work and providing GPU related help.

\bibliographystyle{elsarticle-num}
\bibliography{01_batch_kron}

\begin{thebibliography}{10}
\expandafter\ifx\csname url\endcsname\relax
  \def\url#1{\texttt{#1}}\fi
\expandafter\ifx\csname urlprefix\endcsname\relax\def\urlprefix{URL }\fi
\expandafter\ifx\csname href\endcsname\relax
  \def\href#1#2{#2} \def\path#1{#1}\fi

\bibitem{tgemm}
C.~Jhurani, P.~Mullowney,
  \href{\url{www.ices.utexas.edu/$\char126$chetan/preprints/2013-CJ-PM-GEMM.pd%
f}}{{A GEMM interface and implementation on NVIDIA GPUs for multiple small
  matrices}}, Submitted.
\newline\urlprefix\url{\url{www.ices.utexas.edu/$\char126$chetan/preprints/201%
3-CJ-PM-GEMM.pdf}}

\bibitem{cublas}
{NVIDIA CUDA Basic Linear Algebra Subroutines (cuBLAS) library},
  \url{https://developer.nvidia.com/cublas}, accessed: Feb 21, 2013.

\bibitem{vanloankron}
C.~F.~V. Loan, {The ubiquitous Kronecker product}, Journal of Computational and
  Applied Mathematics 123~(1--2) (2000) 85 -- 100.

\bibitem{deville2002hom}
M.~Deville, P.~Fischer, E.~Mund, {High-Order Methods for Incompressible Fluid
  Flow}, Cambridge University Press, 2002.

\bibitem{hackbusch-tensor}
W.~Hackbusch, Tensor Spaces and Numerical Tensor Calculus, Springer--Verlag,
  Berlin, 2012.

\bibitem{cite:lyap2}
A.~C. Antoulas, Approximation of Large-Scale Dynamical Systems, Cambridge
  University Press, 2005.

\bibitem{Loan93approximationwith}
C.~V. Loan, N.~Pitsianis, {Approximation with Kronecker Products}, in: {Linear
  Algebra for Large Scale and Real Time Applications}, Kluwer Publications,
  1993, pp. 293--314.

\bibitem{cite:lapack}
E.~Anderson, Z.~Bai, C.~Bischof, S.~Blackford, J.~Demmel, J.~Dongarra,
  J.~Du~Croz, A.~Greenbaum, S.~Hammarling, A.~McKenney, D.~Sorensen, {LAPACK}
  Users' Guide, 3rd Edition, SIAM, Philadelphia, PA, 1999.

\bibitem{fermi_gemm}
R.~Nath, S.~Tomov, J.~Dongarra, An improved magma gemm for fermi graphics
  processing units, Int. J. High Perform. Comput. Appl. 24~(4) (2010) 511--515.
\newblock \href {http://dx.doi.org/10.1177/1094342010385729}
  {\path{doi:10.1177/1094342010385729}}.

\bibitem{spectralHP}
G.~Karniadakis, S.~Sherwin, {Spectral/hp Element Methods for CFD}, Oxford
  University Press, USA, 1999.

\bibitem{cite:hpbook}
L.~Demkowicz, Computing with hp-ADAPTIVE FINITE ELEMENTS: Volume I: One and Two
  Dimensional Elliptic and Maxwell Problems, Chapman \& Hall/CRC Press, 2006.

\bibitem{cite:hpbook2}
L.~Demkowicz, W.~Rachowicz, D.~Pardo, M.~Paszynski, J.~Kurtz, A.~Zdunek,
  Computing with hp-ADAPTIVE FINITE ELEMENTS: Volume II Frontiers: Three
  Dimensional Elliptic and Maxwell Problems with Applications, Chapman \&
  Hall/CRC Press, 2007.

\bibitem{solin2003higher}
P.~{\v{S}}ol{\'\i}n, K.~Segeth, I.~Dole{\v{z}}el, {Higher-order Finite Element
  Methods}, Chapman \& Hall/CRC, 2003.

\bibitem{cite:highamASNA2}
N.~J. Higham, Accuracy and Stability of Numerical Algorithms, 2nd Edition, SIAM
  Books, Philadelphia, 2002.

\bibitem{hornjohnson}
R.~A. Horn, C.~R. Johnson, Matrix Analysis, Cambridge University Press, 1990.

\bibitem{lakos}
J.~Lakos, {Large-Scale C++ Software Design}, {Addison-Wesley} professional
  computing series, {Addison-Wesley} Pub. Co, Reading, Mass, 1996.

\end{thebibliography}

\end{document}